\documentclass[twocolumn,superscriptaddress]{revtex4-1}

\usepackage[pdftex]{graphicx}
\usepackage{gensymb}
\usepackage{braket}

\begin{document}

\title{Nanodiamond-enhanced MRI}

\author{David E. J. Waddington}
\affiliation{A. A. Martinos Center for Biomedical Imaging, 149 Thirteenth St., Charlestown, MA 02129, USA}
\affiliation{ARC Centre of Excellence for Engineered Quantum Systems, School of Physics, University of Sydney, Sydney, NSW 2006, Australia}
\affiliation{Department of Physics, Harvard University, 17 Oxford St., Cambridge, MA 02138, USA}

\author{Mathieu Sarracanie}
\affiliation{A. A. Martinos Center for Biomedical Imaging, 149 Thirteenth St., Charlestown, MA 02129, USA}
\affiliation{Department of Physics, Harvard University, 17 Oxford St., Cambridge, MA 02138, USA}
\affiliation{Harvard Medical School, 25 Shattuck St., Boston, MA 02115, USA}

\author{Huiliang Zhang}
\affiliation{Department of Physics, Harvard University, 17 Oxford St., Cambridge, MA 02138, USA}
\affiliation{Harvard-Smithsonian Center for Astrophysics, 60 Garden St., Cambridge, MA 02138, USA}

\author{Najat Salameh}
\affiliation{A. A. Martinos Center for Biomedical Imaging, 149 Thirteenth St., Charlestown, MA 02129, USA}
\affiliation{Department of Physics, Harvard University, 17 Oxford St., Cambridge, MA 02138, USA}
\affiliation{Harvard Medical School, 25 Shattuck St., Boston, MA 02115, USA}

\author{David R. Glenn}
\affiliation{Department of Physics, Harvard University, 17 Oxford St., Cambridge, MA 02138, USA}
\affiliation{Harvard-Smithsonian Center for Astrophysics, 60 Garden St., Cambridge, MA 02138, USA}

\author{Ewa Rej}
\affiliation{ARC Centre of Excellence for Engineered Quantum Systems, School of Physics, University of Sydney, Sydney, NSW 2006, Australia}

\author{Torsten Gaebel}
\affiliation{ARC Centre of Excellence for Engineered Quantum Systems, School of Physics, University of Sydney, Sydney, NSW 2006, Australia}

\author{Thomas Boele}
\affiliation{ARC Centre of Excellence for Engineered Quantum Systems, School of Physics, University of Sydney, Sydney, NSW 2006, Australia}

\author{Ronald L. Walsworth}
\affiliation{Department of Physics, Harvard University, 17 Oxford St., Cambridge, MA 02138, USA}
\affiliation{Harvard-Smithsonian Center for Astrophysics, 60 Garden St., Cambridge, MA 02138, USA}

\author{David J. Reilly}
\affiliation{ARC Centre of Excellence for Engineered Quantum Systems, School of Physics, University of Sydney, Sydney, NSW 2006, Australia}

\author{Matthew S. Rosen*}
\affiliation{A. A. Martinos Center for Biomedical Imaging, 149 Thirteenth St., Charlestown, MA 02129, USA}
\affiliation{Department of Physics, Harvard University, 17 Oxford St., Cambridge, MA 02138, USA}
\affiliation{Harvard Medical School, 25 Shattuck St., Boston, MA 02115, USA}

\begin{abstract}
Nanodiamonds are of interest as nontoxic substrates for targeted drug delivery and as highly biostable fluorescent markers for cellular tracking.  Beyond optical techniques however, options for noninvasive imaging of nanodiamonds in vivo are severely limited.  Here, we demonstrate that the Overhauser effect, a proton-electron double resonance technique developed to detect free radicals in vivo, can enable high contrast magnetic resonance imaging (MRI) of nanodiamonds in water at room temperature and ultra-low magnetic field.  The technique transfers spin polarization from paramagnetic impurities at nanodiamond surfaces to $^1$H spins in the surrounding water solution, creating MRI contrast on-demand. We further examine the conditions required for maximum enhancement as well as the ultimate sensitivity of the technique. The ability to perform continuous hyperpolarization via the Overhauser mechanism, in combination with excellent in vivo stability, raises the possibility of performing noninvasive tracking of nanodiamonds over indefinitely long periods of time.
\end{abstract}

\maketitle


Nanoparticles are rapidly emerging as powerful theranostic substrates \cite{Min2015} for the targeted delivery of vaccines \cite{Reddy2007}, chemotherapy agents \cite{Gaur2014}, immunotheraputics \cite{Almeida2014} and as a means of tracking tumor distribution on whole-body scales \cite{Liu2016,Harisinghani2003}. Biocompatible nanodiamonds (NDs) are ideal examples, featuring surfaces that are readily functionalized to enable tissue growth and their selective uptake by disease processes \cite{Chow2011,Xi2014,Thalhammer2010, Zhang2011}. Imaging NDs in vivo, however, has been to date, mostly limited to sub-cellular environments that are optically accessible \cite{Schrand2007,Mcguinness2011}. Without imaging modalities beyond optical florescence, realization of the full theranostic potential of ND to track and investigate complex disease processes, such as metastatic disease, is unlikely.

Magnetic resonance imaging (MRI) is the gold standard for noninvasive high-contrast imaging of disease in radiology, but has proven ineffective for directly detecting NDs in vivo due to the low abundance and small gyromagnetic ratio of spin-active $^{13}$C nuclei that comprise the carbon lattice.  Hyperpolarization of the $^{13}$C nuclei at cryogenic temperatures can, in principle, overcome the inherently weak signal from diamond by boosting it some 10,000 times to enable MRI contrast from nanoparticle compounds \cite{Rej2015a,Casabianca2011,Dutta2014}.  Despite these prospects, hyperpolarized nuclei always relax to their thermal polarization in a time that, for the smaller sub-micron particles, is currently short enough to limit the usefulness of the method in an imaging context \cite{Cassidy2013,Rej2015a}. 

An alternative approach to tracking concentrations of ND involves functionalizing the ND surface with paramagnetic Gd(III)-chelates to create complexes that could potentially be imaged in vivo with conventional $T_1$-weighted MRI \cite{Manus2010}.  However, this approach faces the challenges of a large background signal and concerns over the potential toxicity of gadolinium-based compounds in the long term \cite{McDonald2015}.

Here, we demonstrate a viable means of imaging and tracking water-ND solutions in vivo using Overhauser magnetic resonance imaging (OMRI) \cite{Lurie2002,Golman2002,Koonjoo2014,Ichikawa2012}.  Operation at ultra-low magnetic field (ULF) enables efficient and biocompatible Overhauser polarization transfer between electronic and nuclear spins.  RF pulsing of the electron paramagnetic resonance (EPR) transition between MRI signal acquisitions continually transfers spin polarization from the paramagnetic centers at the surface of ND to $^1$H nuclei in the surrounding water \cite{Sarracanie2013a}. The presence of ND in the solution thus leads to an enhancement in the $^1$H MRI signal that can readily produce images with contrast sensitive to ND concentrations. Our approach overcomes the limitations imposed by short spin relaxation times of smaller particles and enables switchable tracking of ND solutions over indefinite timescales. In addition to producing images to demonstrate the approach, we investigate the conditions under which this technique leads to maximum sensitivity to the presence of ND, presenting data characterising the efficiency of the Overhauser mechanism as a function of particle concentration and size. These results significantly enhance the theranostic capabilities of non-toxic, bio-functionalized ND, opening the possibility that MRI can be used to monitor and track ND-compounds in vivo.

\section*{The Overhauser effect in nanodiamond solutions}

Various types of ND were used in this study, including high-pressure, high-temperature (HPHT), natural (NAT) and detonation (DET) NDs in sizes from 4 nm - 125 nm.  We focus on results obtained from HPHT 18 nm and HPHT 125 nm NDs as typical representatives of the general behaviour observed.  An air oxidization process, known to etch the ND surface, produces additional variants of NDs for comparison with the commercially sourced varieties \cite{Gaebel2012}.  Aqueous solutions of ND in deionized (DI) water were prepared using high power probe sonication, with HPHT NDs exhibiting the most stability in solution.  HPHT 125 nm solutions show no aggregation over a period of months and a zeta potential of -55 mV [see methods for further details on ND preparation].

\begin{figure}[h]
	
	\includegraphics[width =88mm]{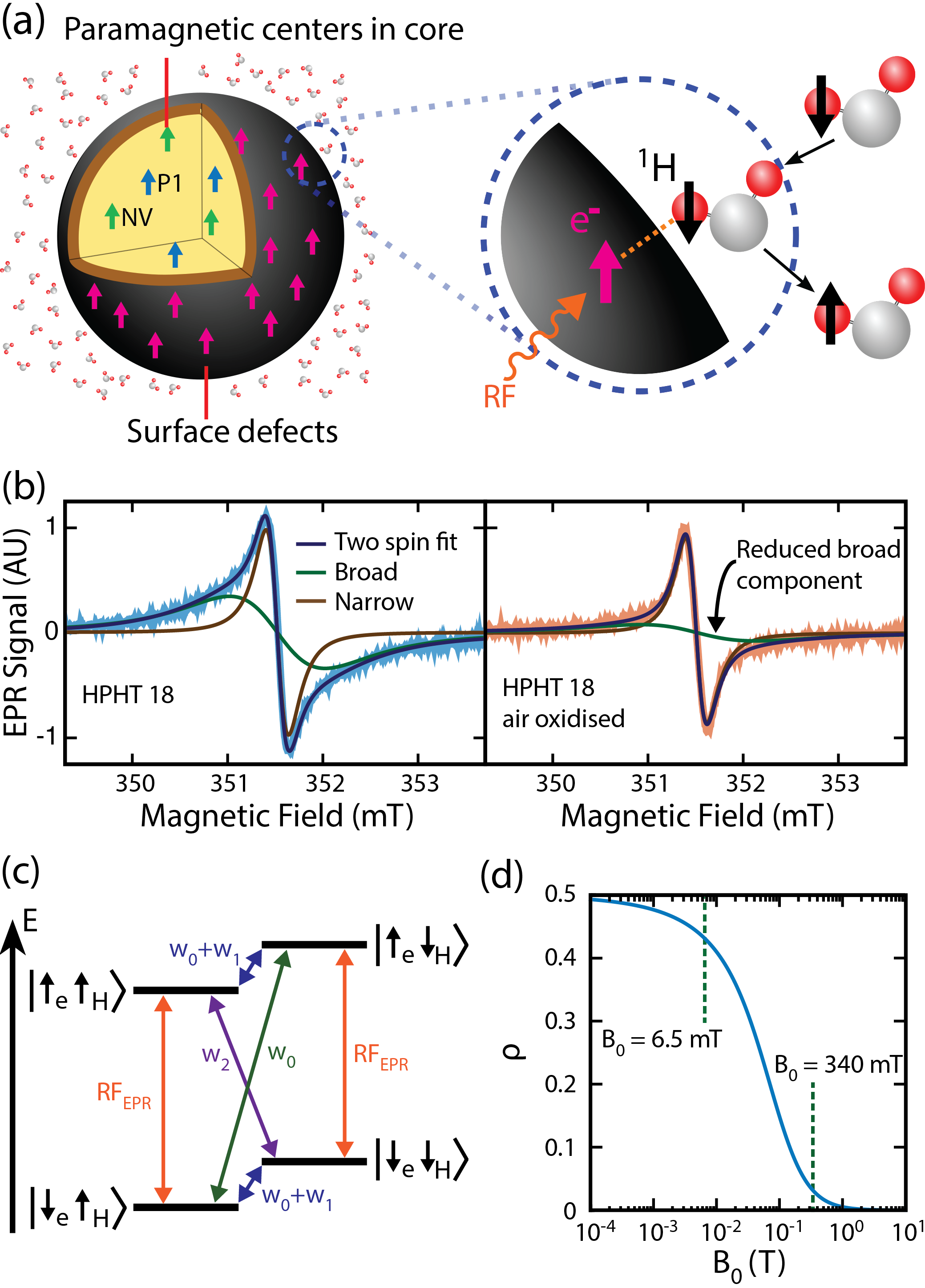}	
	
	\caption{\label{fig1} (a) Schematic of the Overhauser effect at the ND-water interface. (b) X-band EPR spectra of NDs in 100 mg$\cdot$mL$^{-1}$ solutions of DI water.   The left panel shows the spectrum of HPHT 18 nm NDs (blue) and the right panel shows the spectrum of the same HPHT 18 nm NDs after air oxidation (orange).  Both EPR spectra are fitted with a two-spin model (dark blue) that is the sum of a broad spin-1/2 component (green) and a narrow spin-1/2 component (brown).  Air oxidation is seen to remove the broad component whilst leaving the narrow component relatively unaffected. (c) Zeeman split electron and nuclear spin levels in a magnetic field.  Zero-quantum (w$_0$), single-quantum (w$_1$) and double-quantum transitions (w$_2$) are shown.  If the w$_2$ transition dominates, when the EPR transition is pumped, there is a net movement to the $\Ket{\downarrow_e \downarrow_{^1 \text{H}}}$ state.   (d) Coupling factor $\rho$ as a function of magnetic field for a translational correlation time ($\tau_c$) of 430 ns.}

\end{figure}
The basis for detecting and imaging ND in solution is shown in Fig. \ref{fig1}(a).  Image contrast arises from the Overhauser effect, which as a starting point, requires a reservoir of partially polarized electron spins \cite{Ravera2016}. Driving these electrons with a resonant ac magnetic field transfers spin polarization to the interacting $^1$H nuclei in the surrounding solution, (see Fig. \ref{fig1}(a))  \cite{Clarkson1998, Ardenkjaer-Larsen1998}. NDs provide such a reservoir in the form of paramagnetic impurities such as nitrogen vacancy centers, substitutional nitrogen (P1) centers, and unpaired electrons at the nanoparticle surface \cite{Panich2015,Cui2014}. We first characterize our NDs using EPR spectroscopy, determining their impurity content and suitability for Overhauser imaging. 

The EPR spectra of our HPHT 18 nm ND solution is shown in Fig. \ref{fig1}(b) and fits a two-component spin-1/2 model comprising a broad (1.2 mT) component and a narrow (0.2 mT) component (solid lines in figure) \cite{Stoll2006}.  Air oxidation of NDs reduces the amplitude of the broad component in the spectra, presumably by removing the paramagnetic centers at the surface.  Our results are consistent with previous studies suggesting that the broad component is due to disordered dangling bonds at the surface of the ND with the narrow component arising from lattice defects in the crystalline core \cite{Yavkin2015}.  Other types of ND studied here demonstrate similar spectral components [see \emph{SI} for further details].

Having established that ND provides a paramagnetic reservoir suitable for the Overhauser effect, we turn now to address the additional conditions that must be satisfied to enable imaging. Given the ensemble average of the z-component of the nuclear spin over the nuclear moment, the enhancement, $\epsilon = \braket{I_z}/I_0$, generated by the Overhauser mechanism is a function of four parameters \cite{Gunther2011}:

\begin{equation}
  \epsilon = 1 - \rho f s \frac{\left| \gamma_e \right|}{\gamma_n}
  \label{eqn:enhancement}
\end{equation}
where $\rho$ is the coupling factor between electron and nuclear spins, $f$ the leakage factor, $s$ the saturation factor, and $\gamma_e$ and $\gamma_n$ are the electron and nuclear gyromagnetic ratios. Addressing first the coupling factor $\rho$, we note that, when there is dipolar coupling but no hyperfine contact interaction between spins, $\rho$ takes a positive value determined by the correlation time of the two spins, diffusion coefficients and the EPR frequency.  Relatively long correlation times are expected at ND surfaces due to the formation of a nanophase of water with 1 nm thickness at the ND-water interface \cite{Korobov2007}.  Assuming free diffusion of water at a distance of 1 nm from the ND surface, we follow refs. \cite{Armstrong2009,Toth2002,Hwang1975,Freed1978} and plot the field dependence of $\rho$ for a calculated correlation time of 430 ns, as shown in Fig. \ref{fig1}(d) [see \emph{SI} for details of calculation]. Not surprisingly, given the long correlation time between spins, increasing the magnetic field above a few milli-Tesla rapidly suppresses the mutual flip-flip of dipolar coupled electron and nuclear spins and thus the nuclear enhancement possible via the Overhauser effect. Our choice of magnetic field for Overhauser imaging is thus constrained to the ULF regime, where serendipitously, the frequency of the EPR field produces minimal heating from dielectric loss associated with water at $20\degree$C \cite{Gabriel1998}.

\begin{figure}[h]
	
	\centering
	\includegraphics[width=88mm]{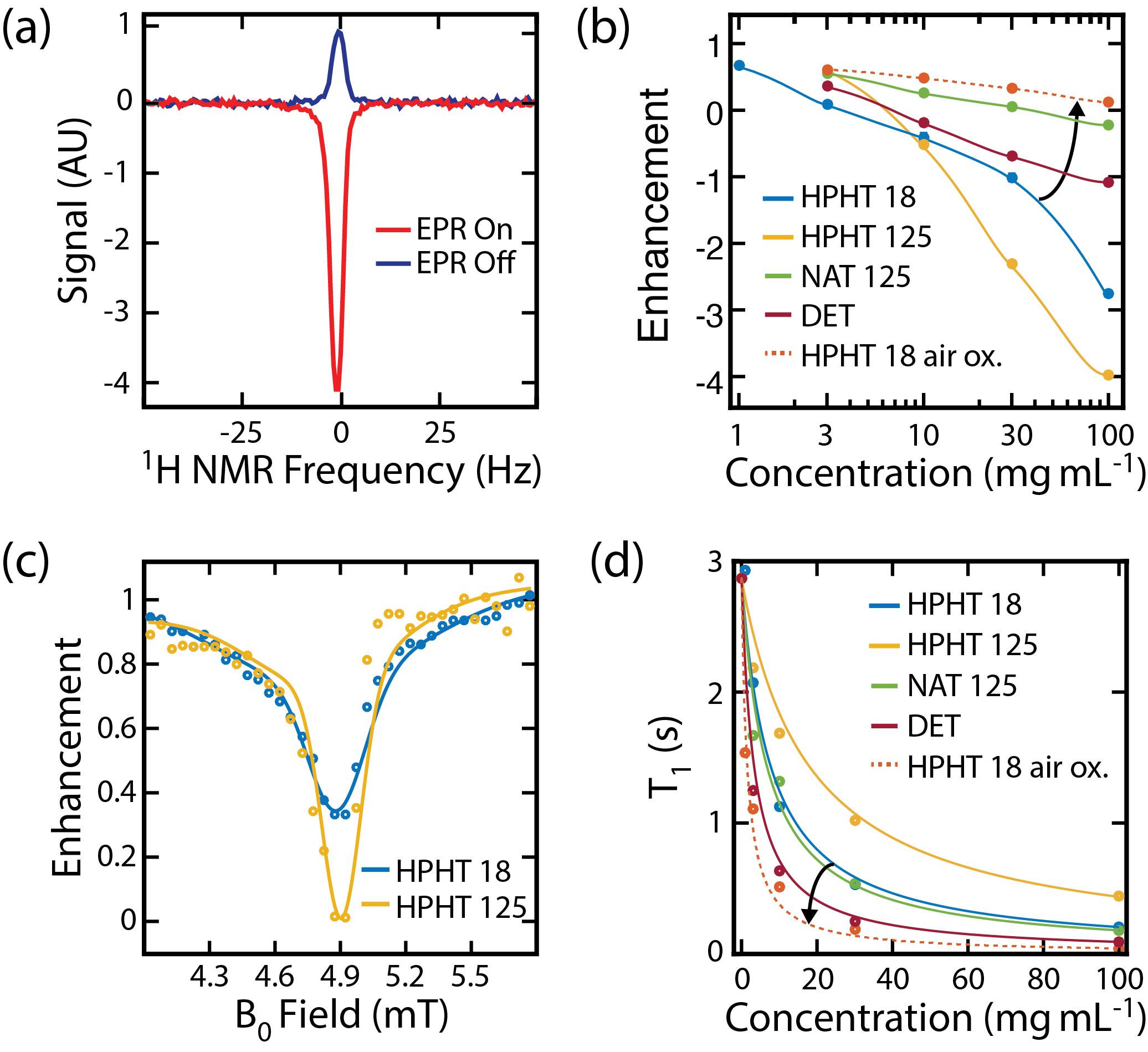}
	\caption{(a) Overhauser enhancement of $^1$H polarization in an HPHT 125 nm 100 mg$\cdot$mL$^{-1}$ ND solution.  The DNP $^1$H spectrum (red) is enhanced by -4.0 over the thermal $^1$H spectrum (blue).   The enhanced spectrum was acquired after the EPR transition had been driven for 1.5 s with 49 W of RF power at 190 MHz. (b) $^1$H saturation enhancement vs concentration for ND solutions at 49 W, 6.5 mT and 190 MHz. The Overhauser effect was observed for HPHT (blue - 18 nm, yellow - 125 nm), NAT 125 nm (green), DET (red) and air oxidized HPHT 18 nm NDs (orange).  Solid lines are included as a guide to the eye.  Arrow indicates the change in enhancement after air oxidation.  (c) $^1$H enhancement vs $B_0$ with EPR pumping at a constant frequency of 140 MHz with 24 W of power.  $^1$H NMR detection was performed on resonance.  Aqueous solutions of HPHT 18 nm (blue) and HPHT 125 nm (yellow) at 50 mg$\cdot$mL$^{-1}$ concentration were used.  Solid lines are included as a guide to the eye. (d) $T_{1}$ relaxation times of ND solutions at 6.5 mT.  Solid lines are a fit to the concentration dependent relaxivity equation.  Arrow indicates the change in relaxivity after air oxidation.  The $T_1$ relaxivity coefficients are 
	4.5$\times 10^{-2}$ mL$\cdot$s$^{-1} \cdot$mg$^{-1}$ for HPHT 18 nm (blue), 
	1.9$\times 10^{-2}$ mL$\cdot$s$^{-1} \cdot$mg$^{-1}$ for HPHT 125 nm (yellow), 
	5.2$\times 10^{-2}$ mL$\cdot$s$^{-1} \cdot$mg$^{-1}$ for NAT 125 nm (green), 
	1.0$\times 10^{-1}$ mL$\cdot$s$^{-1} \cdot$mg$^{-1}$ for DET (red) and 
	2.3$\times 10^{-1}$ mL$\cdot$s$^{-1} \cdot$mg$^{-1}$ for air oxidized HPHT 18 nm (orange).}
	\label{fig:physics}
\end{figure}

To demonstrate that NDs can be detected via the Overhauser effect at ULF, we set $B_0$ = 6.5 mT and apply an RF magnetic field at the EPR frequency of 190 MHz to a HPHT 125 nm, 100 mg$\cdot$mL$^{-1}$ sample.  The $^1$H signal from the water surrounding the ND is then detected through standard inductive NMR techniques after the $^1$H system has reached equilibrium [see methods]. Under these conditions we observe an enhancement of -4.0 in the $^1$H spin polarization when EPR power is applied, as shown in Fig. \ref{fig:physics}(a).

Examining the enhancement produced by different types of ND, Fig. \ref{fig:physics}(b) shows the sensitivity of the technique to nanoparticle concentration. We draw attention to the data for the HPHT 18 nm NDs, which indicates that at concentrations of 1 mg$\cdot$mL$^{-1}$, a 33 \% change in polarization can be observed.  Natural NDs produce a relatively small enhancement, probably due to a relatively low concentration of paramagnetic defects, as seen in their EPR spectra [see Fig. S1(c) for spectra].

Having demonstrated that Overhauser enhancement is possible with ND, we return to Eq. \ref{eqn:enhancement} to further consider the conditions needed for optimal imaging. The saturation factor $s$ describes the proportion of the EPR linewidth that is driven and takes a maximum value of 1 when the electron transitions are completely saturated at high RF power.  To measure the EPR linewidth at ULF we sweep $B_0$  while driving electron transitions at 140 MHz.   As shown in Fig. \ref{fig:physics}(c), HPHT 18 nm and HPHT 125 nm solutions show a linewidth for the enhancement of $\sim$ 0.3 mT at a frequency consistent with the gyromagnetic ratio of a free electron.  This result indicates that the paramagnetic centers responsible for the Overhauser effect can be saturated with an ac field, $B_{1e}$, of 1 mT, which is easily achieved in our spectroscopic probe \cite{Mispelter2006}.

Finally, the remaining parameter in Eq. \ref{eqn:enhancement} is the leakage factor $f$ which describes how effectively electrons relax the nuclear spin environment, taking a maximum value of 1 when all nuclear spin relaxation is via paramagnetic centers.  Measuring the spin relaxation time $T_{1}$ of $^1$H spins in ND solutions (without applying EPR power) provides a means of characterizing $f$ in these systems, as shown in Fig 2(d).  We note that ND solutions with shorter $T_1$, and hence larger $f$, do not necessarily give a higher Overhauser enhancement.  For example, NAT 125 nm NDs show a $T_1$ relaxivity more than double that of HPHT 125 nm NDs, despite showing a much smaller enhancement in Fig. \ref{fig:physics}(b).  Presumably, $\rho$ is suppressed in the quasistatic nanophase by slow diffusion of water molecules and the increased paramagnetic nuclear relaxation rate \cite{Rej2016}.  $^1$H nuclei `trapped' in the nanophase will experience rapid spin-lattice relaxation, giving the $f$ we observe and an overall enhancement that depends on the specifics of each ND surface.

Solutions prepared with air oxidized ND  consistently exhibit reduced enhancements and higher $T_1$ relaxivity, as shown for 18 nm HPHT air oxidized NDs in Figs. 2(b) and 2(d).  The increased nuclear spin-lattice relaxation rate will contribute to a reduction in Overhauser enhancement and we speculate that the enhancement is further reduced due to a lower concentration of paramagnetic centers after removal of surface impurities by air oxidation.

\section*{Overhauser MRI with Nanodiamond}

\begin{figure}[h]
	\centering
		\includegraphics[width =88mm]{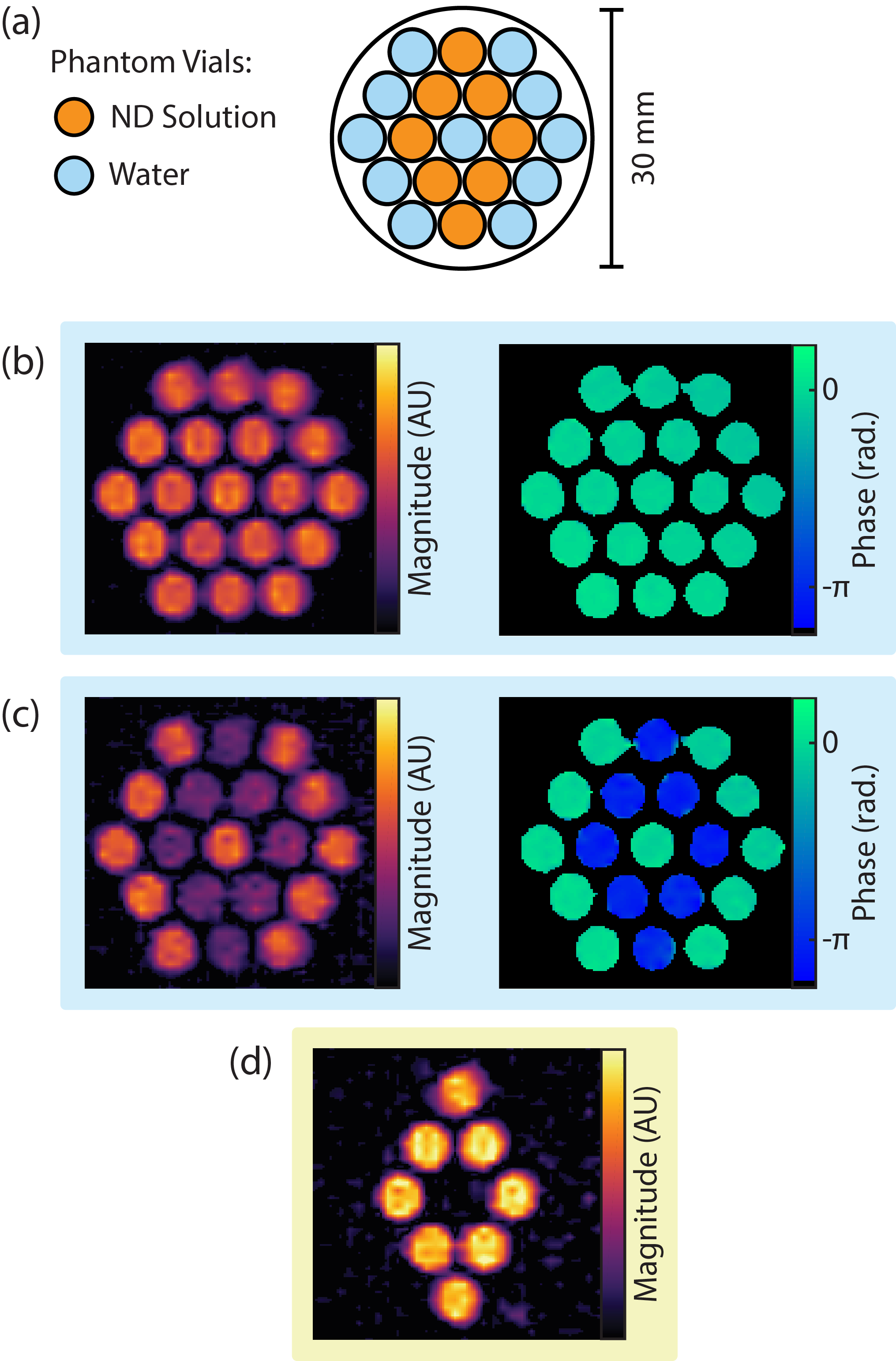}

\caption{ND imaging with Overhauser MRI (OMRI) at 6.5 mT. (a) Imaging phantom.  Vials of DI water (blue) and vials of HPHT 125 nm ND at 100 mg$\cdot$mL$^{-1}$ (orange) were arranged in the pattern shown. (b) Standard \emph{b}-SSFP MRI of the phantom shown in a. (c) OMRI \emph{b}-SSFP image of the same phantom.  The Overhauser effect generates contrast between ND solution and water.  The phase of the acquired signal is inverted in the ND solution. (d) The difference of MRI and OMRI acquisitions.  The background signal is suppressed, showing signal only where ND is present.}
\label{fig:images}
\end{figure}

With the conditions that lead to near optimal Overhauser enhancement now established, we turn to demonstrate that this approach can indeed be used as the basis for detecting ND solutions using ultra-low field MRI. Imaging is performed using a custom proton-electron, double resonant probe in an open-access, low-field, human MRI scanner operating at a $B_0$ of 6.5 mT \cite{Sarracanie2015}.  To display the MRI contrast possible between a ND solution and water we make use of the phantom illustrated in Fig. \ref{fig:images}(a), which consists of glass vials filled with 500 $\mu$L of either DI water or aqueous solutions of HPHT 125 nm ND at 100 mg$\cdot$mL$^{-1}$ and organized in a diamond-shaped pattern.

To overcome the low sensitivity of MRI at ultra-low field, the phantom is imaged with a conventional 2D \emph{b}-SSFP (balanced steady state free precession) MRI sequence in which $1/3$ of the imaging time is spent acquiring signal [see methods for details] \cite{Scheffler2003}.  Although good spatial resolution is achieved, no discernible contrast is evident between ND solution vials and water vials, as shown in Fig. \ref{fig:images}(b). This is not surprising given that contrast using the \emph{b}-SSFP sequence is produced via $^1$H concentration and the ratio $T_2/T_1$, which are approximately equal for all vials in the phantom [see Fig. \ref{fig:physics}(d) for $T_1$ and Fig. S1(b) for $T_2$].  We note that obtaining contrast with \emph{b}-SSFP at ULF is usually challenging, as when $B_0 \rightarrow 0$, it is a general result that $T_2/T_1 \rightarrow 1$ \cite{Bottomley1984}.

The phantom is then imaged with a 2D OMRI \emph{b}-SSFP sequence, as shown in Fig. \ref{fig:images}(c).  The OMRI \emph{b}-SSFP sequence is equivalent to the regular \emph{b-}-SSFP sequence except the EPR transition of the ND solution is driven during the phase encode period \cite{Sarracanie2013a}.  The appearance of water vials in the OMRI \emph{b}-SSFP image is unchanged from the regular \emph{b}-SSFP image.  However, the ND solutions demonstrate significant relative contrast, with a change in magnitude and inversion of signal phase, as a result of the negative enhancement from the Overhauser effect.

The switchable nature of the Overhauser contrast allows us to take the difference of the signal in MRI and OMRI images to generate the image in Fig. \ref{fig:images}(d).  Such a difference image suppresses the background signal, clearly showing the spatial distribution of NDs.  For this difference image we calculate a contrast to noise ratio (CNR) of $\sim$ 30, with the OMRI scan comprising 80 averages, taking 3 min 40 s to complete.

Having shown the feasilibity of ND imaging with OMRI, we can derive the concentration sensitivity limit of the technique from the magnitude of the Overhauser effect and the MRI sensitivity in our setup.  Images in Fig. \ref{fig:images} were acquired with 1.5 mg ND per pixel.  Assuming a minimum detectable CNR of 2, the magnitude of the Overhauser effect must be 7\% of that used in Fig.  \ref{fig:images}.  Our measurements indicate that, under this criteria, HPHT 18 nm NDs should be still detectable at concentrations of 1 mg$\cdot$mL$^{-1}$ (see Fig. \ref{fig:physics}(a)).  This sets a lower limit for detection of 15 $\mu$g of ND per pixel, or a particle molar sensitivity of 150 nM.  We note that this particle mass sensitivity is equivalent to that reported for other hyperpolarized MRI particle imaging modalities \cite{Cassidy2013}.

\section*{Discussion and Conclusion}

NDs are non-toxic at high concentrations and resist in vivo degradation for periods of over a month \cite{Vaijayanthimala2012}.  Thus, the results presented here illustrate the potential of nanodiamond OMRI as a practical methodology for long-term biological imaging providing new types of contrast and functionality.  All imaging was performed on systems designed for in vivo OMRI with RF powers acceptable for use in vivo \cite{Massot2012}, raising the possibility of several exciting biological applications, for example, in the accurate detection of lymph node tumors, which is vital to the treatment of metastatic prostate cancer \cite{Harisinghani2003}.  Given that nanoparticles $\sim$ 25 nm in size are known to preferentially accumulate only in healthy lymphatic tissue, the ability to detect and image ND in vivo may enable isolation of the disease \cite{Reddy2007}. OMRI using NDs of selective size ranges could also allow targeted probing of the permeability of the blood brain barrier, giving a quantifiable measure of disruption in the acute stage of ischemic stroke \cite{Roney2005}. 

Finally, we remark that surface functionalization could be used  to tailor the Overhauser contribution produced by the ND. In this way therapeutic agents attached to the surface could suppress the Overhauser effect by increasing the distance between radicals at their surface and free water, leaving them `dark' in OMRI scans.  After targeted drug release, the Overhauser effect could return to normal, showing up `bright' with OMRI and enabling effective tracking of the site of drug delivery. The OMRI approach may also enable the hyperpolarization of fluids flowing across the surface of diamond nanostructures \cite{Lingwood2012,Abrams2014,Meriles2014}.

In conclusion, we have used recent advances in ULF MRI to extend the usefulness of OMRI to nanoparticle imaging.  The ability to noninvasively image biocompatible NDs with switchable contrast at biologically relevant concentrations is promising for a range of diagnostic applications.  Switchable contrast allows suppression of the background signal present in other $T_1$ and $T_2$ based nanoparticle MRI modalities \cite{Bouchard2009,Manus2010,Xing2009}.  Furthermore, the long term biological stability of nanodiamonds in vivo, as well as the unlimited repeatability of the hyperpolarization sequence, raises the possibility of imaging metabolic processes over dramatically longer timescales than is possible with ex situ hyperpolarization techniques \cite{Cassidy2013,Rej2015a}.

\section*{Materials and Methods}

\textbf{Spectroscopic measurements at 6.5 mT.}  The $^1$H enhancement of ND solutions was measured by saturating the EPR transition at 190 MHz for a period of $5T_1$ using 31 W of RF power before a standard NMR FID acquisition.  The magnitude of the hyperpolarized FID is compared to an FID taken at thermal equilibrium.  A high filling factor Alderman-Grant Resonator was used for EPR with an orthogonal solenoid used for NMR detection.  $T_1$ relaxation times were measured using a conventional inversion recovery acquisition.

\textbf{Overhauser MRI.}  Imaging was performed at 6.5 mT in our ultra-low field MRI scanner \cite{Sarracanie2015} using a \emph{b}-SSFP Overhauser-enhanced MRI (OMRI) sequence at room temperature \cite{Sarracanie2013a}.  A homebuilt imaging probe was constructed from an Alderman-Grant resonator (EPR: 190 MHz) and a solenoid ($^1$H: 276 kHz).  The EPR resonator was pulsed on during the phase encode steps, with 62 W delivered to the EPR resonator at a duty cycle of 52\%.  Gradient strength was a maximum of 1 mT$\cdot$m$^{-1}$.  Images in Fig. \ref{fig:images} were acquired with a 256 x 40 matrix size and cropped. Data was acquired in 2D with a pixel size of 1.0 mm x 0.76 mm over a 30 mm x 30 mm field of view and interpolated onto 0.25 mm x 0.19 mm pixels.  The phantom thickness was 20 mm.  The standard MRI was acquired with 200 averages (9 minutes) and the OMRI image with 80 averages (3 minutes 40 seconds).  Image magnitude was scaled for an accurate comparison.

\textbf{Nanodiamond Solution Preparation.}  NDs used in this study were sourced from Microdiamant, Switzerland.  Nanodiamond types used were; monocrystalline, synthetic HPHT NDs in 18 nm (0-30 nm, median diameter 18 nm) and 125 nm (0-250 nm, median diameter 125 nm) sizes; monocrystalline, 125 nm NAT NDs (0-250 nm, median diameter 125 nm); and polycrystalline DET ND (Cluster size 250-1000 nm, median 500 nm.  Individual particle size 4-8 nm).  Size specifications were provided by the manufacturer.  Air oxidized NDs were prepared by placing them in a furnace at standard pressure for one hour at 550 {\degree}C after an initial temperature ramp \cite{Gaebel2012}.  ND samples were mixed with DI water and sonicated with a Branson probe sonicator at 120 W and 50\% duty cycle for 40 minutes to disaggregate ND clusters.  

Particle size and zeta potential measurements were performed on ND solutions in a Beckman Coulter Delsa Nano C Particle Analyzer.  Particle size measurements confirmed that monocrystalline NDs were well dispersed in water after sonication.  Particle sizes  of monocrystalline NDs were found to be consistent with manufacturer specifications.  DET NDs still displayed some clustering and inconsistent particle size in solution after probe sonication.

\textbf{DNP Linewidth Measurements - \emph{B}$_0$ Field Sweep.}  The linewidth of the Overhauser effect was studied by measuring the $^1$H enhancement at various magnetic field strengths whilst driving the EPR resonator at a frequency of 140 MHz and 25 W.    The EPR frequency was lowered from 190 MHz in order to capture the enhancement profile either side of the peak without exceeding the maximum field accessible in our ULF magnet.  $^1$H enhancement was measured as the magnetic field was stepped between 4 mT and 7 mT.  NMR detection was performed at the $^1$H resonance for any given field strength, with a low $Q$ solenoidal coil.

\textbf{EPR Measurements.}  CW EPR measurements were performed on 100 mg$\cdot$mL$^{-1}$ ND samples in a Bruker ElexSys E500 EPR system.  Modulation frequency was 100 kHz at an amplitude of 0.1 G and incident microwave power of 0.6725 mW.  Sample volumes in the cavity were kept consistent to allow comparison of relative peak heights.  Individual EPR components were simulated in EasySpin \cite{Stoll2006} and a least squares analysis was used to find the best fit to the data while varying $g$-factor, linewidth and amplitude.

\section*{Acknowledgements}
We thank B.D. Armstrong for valuable discussions. This work was supported by the Australian Research Council Centre of Excellence Scheme (Grant No. EQuS CE110001013) ARC DP1094439, the Lockheed Martin Corporation, U.S. Army Medical Research and Materiel Command (USAMRMC), Defense Medical Research and Development Program (DMRDP) award W81XWH-11-2-0076 (DM09094)
and the U.S. Army Research Office (ARO) Multidisciplinary University Research Initiative (MURI) award W911NF-15-1-0548. D.E.J.W. was supported by ANSTO and the Australian-American Fulbright Commison.  N.S. was supported by the Swiss National Science Foundation (P300P2\_147768).

\section*{Author Contributions}
D.E.J.W., M.S., N.S. and M.S.R. performed the experiment.  D.E.J.W., H.Z., E.R., T.G. and T.B. prepared samples and undertook sample characterization.  H.Z. and D.R.G. performed the EPR experiments.  All authors interpreted, processed and analysed the data.  All authors contributed to writing the manuscript.

*Correspondence and requests for materials should be addressed to M.S.R (email: mrosen@cfa.harvard.edu).

\bibliographystyle{naturemag}
\bibliography{ND_OMRI_arxiv}

\end{document}